\documentclass[%
 reprint,
 superscriptaddress,
 amsmath,amssymb,
 aps,
 pra,
]{revtex4-2}

\usepackage{graphicx}
\usepackage{dcolumn}
\usepackage{bm}
\usepackage[colorlinks=true, allcolors=blue]{hyperref}

\begin{document}

\preprint{APS/123-QED}

\title{Split-Step Dirac Cellular Automata for Continuous-Time Dirac Dynamics on Finite Spatial Lattices}

\author{Wei-Ting Wang}
 \email{d11245002@ntu.edu.tw}
 \email{aroe01325@gmail.com}
 \affiliation{Department of Physics, National Taiwan University, Taipei 106319, Taiwan}

\author{Pei-Ming Ho}
 \email{pmho@phys.ntu.edu.tw}
 \affiliation{Department of Physics, National Taiwan University, Taipei 106319, Taiwan}
 \affiliation{Center for Theoretical Physics, National Taiwan University, Taipei 106319, Taiwan}
 \affiliation{Physics Division, National Center for Theoretical Sciences, Taipei 106319, Taiwan}

\author{Ching-Ray Chang}
 \email{crchang@phys.ntu.edu.tw}
 \affiliation{Department of Physics, National Taiwan University, Taipei 106319, Taiwan}
 \affiliation{Department of Physics and Quantum Information Center, Chung Yuan Christian University, Taoyuan 320314, Taiwan}

\date{\today}

\begin{abstract}
Dirac Cellular Automata (DCA) provide a framework for simulating Dirac dynamics, yet the rigid coupling between spatial and temporal resolutions can introduce artificial phase-matching symmetries on finite grids that suppress interference phenomena such as Zitterbewegung. In this work, we propose a Split-Step Dirac Cellular Automaton (SDCA) that enables continuous-time Dirac evolution at fixed spatial discretization. By employing a Trotterized fractional-step scheme in the momentum representation, SDCA breaks the phase-matching cancellation present in the standard DCA and recovers the interference dynamics of the continuous-time limit. We benchmark the SDCA through analytical and numerical studies and demonstrate its implementation on IBM Quantum processors. Despite the increased circuit depth required for finer temporal resolution, the NISQ implementation reproduces the characteristic velocity oscillations and entanglement-entropy dynamics of the continuous-time model. We further investigate hardware-topology trade-offs and dynamic circuit implementations of the Quantum Fourier Transform (QFT), highlighting the competing effects of gate errors, measurement, and feed-forward latency. These results demonstrate that SDCA provides a practical framework for improving temporal resolution while maintaining a fixed spatial quantum register, enabling the exploration of relativistic quantum dynamics on near-term quantum devices.
\end{abstract}

\maketitle


\section{\label{sec:introduction}Introduction}

Quantum cellular automata (QCA) \cite{Watrous1995,Arrighi2019} were proposed as an extension of classical cellular automata (CA) \cite{Wolfram1984Nature, Wolfram1984Computation} in the quantum domain, offering a discrete and local model for quantum evolution governed by unitary rules. The concept was originally introduced by Feynman \cite{feynman2010quantum} and later formalized in greater mathematical detail by Grössing and Zeilinger \cite{grossing1988quantum}. Since then, QCAs have become tools for simulating quantum dynamics, particularly those involving relativistic particles. Relativistic particles exhibit unique characteristics such as spin, chirality, and the existence of antiparticles \cite{pal2011dirac}. Therefore, simulating the dynamics of relativistic particles is an important topic.

Unlike continuous models, QCAs operate over discrete spacetime grids, making them well-suited for digital implementation on quantum hardware, serving as a bridge between quantum physics and quantum information science \cite{fillion2017algorithm}. To date, several QCA models have been proposed that can recover the Dirac dynamics, such as discrete-time quantum walks (DTQWs) \cite{strauch2006relativistic,bracken2007free,chandrashekar2010relationship} and Dirac cellular automata (DCA) \cite{bialynicki1994weyl,meyer1996quantum,Bisio2015quantum}. They recover the Dirac dynamics with first-order and second-order approximations, respectively \cite{wang2024observing}. These models reproduce relativistic effects like Zitterbewegung and Klein tunneling on quantum computers \cite{kurzynski2008relativistic,bisio2013dirac}. In addition, these models are also applied in studying phenomena such as Anderson localization \cite{schreiber2011decoherence,crespi2013anderson,derevyanko2018anderson}, topological phases \cite{kitagawa2010exploring,flurin2017observing}, and neutrino oscillations \cite{di2016quantum,mallick2017neutrino}, showcasing their versatility in simulating a wide range of quantum effects.

Beyond exhibiting dynamics analogous to the Dirac equation, QCA serves as a prominent platform for investigating quantum information theoretic properties, including von Neumann entropy \cite{mallick2016dirac} and Stabilizer Rényi Entropy (SRE) \cite{leone2022stabilizer,mittal2025quantum}. However, implementing QCAs directly on quantum computers remains a significant challenge. The time evolution of many QCA models requires Multi-Controlled gate operations \cite{huerta2020quantum, singh2021quantum}, resulting in deep quantum circuits. These complexities pose practical difficulties for implementation on current noisy intermediate-scale quantum (NISQ) devices \cite{preskill2018quantum}, where circuit depth is constrained by limited coherence times and gate fidelities.

Recent work on DTQWs has shown that these limitations can be mitigated using the Quantum Fourier Transform (QFT). In the momentum representation, shift operations can be realized by parallel single-qubit phase gates, which drastically reduces circuit complexity. For instance, Saxena et al. demonstrated that QFT-based implementations of DTQWs can be executed with shallower circuits, making them suitable for current near-term quantum hardware \cite{shakeel2020efficient}. In addition, if the QFT is immediately followed by measurement, the $n$-qubit QFT can be efficiently implemented using $\mathcal{O}(n)$ mid-circuit measurements (MCMs) and classical feed-forward (FF) phase gates without any strict qubit connectivity constraints \cite{baumer2024quantum}. In parallel, recent advances in adaptive-step Trotterization schemes \cite{zhao2023making} highlight the benefits of using tunable time steps in quantum simulation. These studies show that adaptively adjusting the time resolution can improve simulation fidelity, even in the presence of hardware noise. Therefore, employing smaller, tunable time steps allows for more accurate tracking of a system’s true continuous-time evolution and enables the partial correction of errors that typically arise from finite-step approximations.

In this study, we extend the standard DCA to the Split-Step Dirac Cellular Automaton (SDCA), formulated in the momentum representation. In this representation, the DCA time-evolution operator can be decomposed into multiple fractional sub-steps. This temporal refinement reduces Trotterization errors and, more importantly, breaks the phase-matching condition imposed by the rigid coupling between spatial and temporal resolutions in the standard DCA, thereby restoring the dependence of the interference dynamics on the relative phase of the initial spinor. In addition, the time evolution can be efficiently implemented in the momentum representation using the QFT on a quantum computer. We therefore not only theoretically analyze the effects of fractional-step evolution but also experimentally demonstrate the SDCA on the IBM Quantum platform \cite{IBMQuantum}, investigating the trade-offs among circuit depth, hardware connectivity, and measurement-based implementations.

The remainder of this paper is organized as follows. Section \ref{sec:theory} reviews the standard DCA and derives the theoretical framework for the SDCA. Section \ref{sec:dynamics} presents the numerical results, analyzing the velocity oscillations and entanglement entropy. Section \ref{sec:NISQ_devices} details the experimental implementation on IBM NISQ devices. Finally, Section \ref{sec:conclusion} summarizes our findings and outlines future perspectives for relativistic quantum simulations.

\begin{figure*}[ht]
\centering
\includegraphics[width=\linewidth]{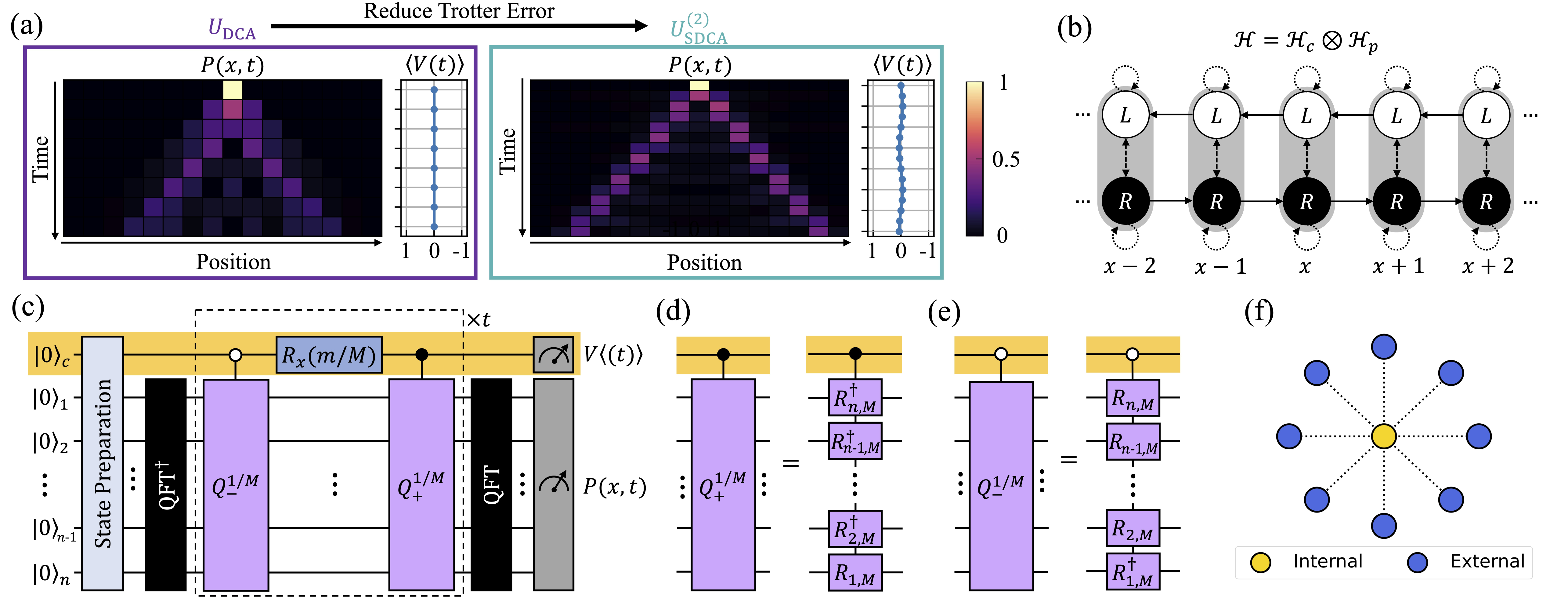}
\caption{
(a) Comparison of DCA and SDCA through spacetime diagrams and velocity expectation values. The initial state of the particle is $|\Psi(t=0)\rangle =\frac{1}{\sqrt{2}}(|0\rangle_{c}+|1\rangle_{c}) \otimes |x=0\rangle$ with mass $m = \pi/4$. 
Left: Evolution under the standard DCA $U_{\mathrm{DCA}}=U_{\mathrm{SDCA}}^{(1)}$. Right: evolution under two sub-step SDCA $U_{\mathrm{SDCA}}^{(2)}$, where the time resolution is refined. (b) Amplitude transition of SDCA in Hilbert space $H = \mathcal{H}_c \otimes \mathcal{H}_p$, where $R$ and $L$ denote the right- and left-handed components of internal space. Solid (dashed) arrows represent the external (internal) transition, and dotted arrows represent the new transition paths (self-loop) emerging from the split-step construction.
(c) The quantum circuit implementation of $U_{\mathrm{SDCA}}^{(M)}$. The external space $\mathcal{H}_p$ is mapped to the external qubits $1,2,\dots,n$ (white background) and the internal space $\mathcal{H}_c$ is mapped into internal qubit $c$ (yellow background). Measurement on the internal qubit yields the expectation value of velocity $\langle V \rangle$; measurement on the position qubits yields the particle’s position distribution $P(x)$. (d)-(e) The quantum circuit of the phase shift operator $Q_{\pm}^{1/M}$, where 
$R_{k,M}=\begin{pmatrix} 1 & 0 \\ 0 & e^{2\pi i/2^{k}M} \end{pmatrix}$. (f) The qubit connectivity used in SDCA in momentum representation}
\label{fig:SDCA_circuit}
\end{figure*}

\section{\label{sec:theory}THEORETICAL FRAMEWORK}
\subsection{\label{sec:DCA}Dirac Cellular Automata}

The dynamics of a (1+1)-dimensional spin-$\frac{1}{2}$ particle with rest mass $m$ can be described by the Dirac Hamiltonian (with $\hbar = 1$, $c = 1$):
\begin{equation}
H = -i \gamma^0 \gamma^1 \partial_x + m \gamma^0,
\label{eq:dirac_ham}
\end{equation}
where $\gamma^\mu$ is a set of 2×2 matrices which satisfy anti-commutation relation $\{\gamma^\mu, \gamma^\nu\} = \gamma^\mu \gamma^\nu + \gamma^\nu \gamma^\mu = 2g^{\mu \nu} I$, ensuring Lorentz invariance. 
The quantum state of such a particle spans a composite Hilbert space $\mathcal{H} = \mathcal{H}_c \otimes \mathcal{H}_p$ where $\mathcal{H}_c$ represents the internal degree of freedom, and $\mathcal{H}_p$ corresponds to the external degree of freedom. The internal space carries information about spin, helicity, or particle-antiparticle identity, while the external space encodes the particle’s position and motion in space. In our formulation, we choose $\gamma^0 \gamma^1 = -\sigma_z$ and $\gamma^0 = \sigma_x$, where $(\sigma_x, \sigma_y, \sigma_z)$ are Pauli matrices. In momentum representation, with $k = -i\partial_x$, Eq.~\eqref{eq:dirac_ham} can be rewritten as: 
\begin{equation}
H = -\sigma_z k + m \sigma_x.
\label{eq:momentum_ham}
\end{equation}

The Dirac cellular automata (DCA) is a discrete-time model that approximates the time evolution governed by this Hamiltonian using a second-order Trotter expansion over a small-time interval $\delta t=1$ and small-space interval $\delta x=c \delta t=1$, hence yielding a approximation error of $\mathcal{O}(\delta t^3)$. The DCA operator in momentum representation is given by:
\begin{align}
U_{\mathrm{DCA}} &= 
\begin{pmatrix} e^{ik} & 0 \\ 0 & I \end{pmatrix}
\begin{pmatrix} \cos(m) & -i \sin(m) \\
-i \sin(m) & \cos(m)
\end{pmatrix}
\begin{pmatrix} I & 0 \\ 0 & e^{-ik} \end{pmatrix} 
\nonumber \\
&=
\begin{pmatrix}
\cos(m) e^{ik} & -i \sin(m) I \\
-i \sin(m) I & \cos(m) e^{-ik}
\end{pmatrix},
\label{eq:u_dca_k}
\end{align}
where $k\in [-\pi,\pi]$.

For an external space of size $N=2^{n}$, where $n$ represents the number of qubits in the external space, the phase shift operator  $e^{\mp ik}$ can be implemented by the diagonal matrix $Q_{\pm}$. Together with the identity matrix $I$, both $Q_{\pm}$ and $I$ are $N\times N$ matrices. 
In the DCA, the shift operator $Q_{-}$ in momentum space is given by
\begin{equation}
Q_- = 
\begin{pmatrix}
1 & 0 & \cdots & 0 & 0 & \cdots & 0 \\
0 & \omega & \cdots & 0 & 0 & \cdots & 0 \\
\vdots & \vdots & \ddots & \vdots & \vdots & \ddots & \vdots \\
0 & 0 & \cdots & \omega^{\frac{N}{2}-1} & 0 & \cdots & 0 \\
0 & 0 & \cdots & 0 & \omega^{-\frac{N}{2}} & \cdots & 0 \\
\vdots & \vdots & \ddots & \vdots & \vdots & \ddots & \vdots \\
0 & 0 & \cdots & 0 & 0 & \cdots & \omega^{-1}
\end{pmatrix}_{N \times N},
\label{eq:Q}
\end{equation}
where $\omega = e^{i\Delta k} = e^{i 2\pi / N}$. Here, the diagonal basis is ordered according to the QFT convention (see Appendix~\ref{APP:A}), where the momentum wrap as $k \in \{0, \Delta k, \dots, \Delta k(N/2-1), -\Delta k(N/2), \dots, -\Delta k\}$.

This operator is unitary and satisfies $Q_-^N = Q_+^N = I$, with $Q_+ = Q_-^\dagger$. The value $\Delta k$ represents the momentum quantum resulting from the discretization of the spatial lattice, and consequently, $\omega$ acts as the fundamental phase shift associated with a unit spatial translation.

By applying the QFT, the DCA operator can be transformed into position representation:
\begin{equation}
U_{\mathrm{DCA}} = 
\begin{pmatrix}
\cos(m) \, T_- & -i \sin(m) \, I \\
-i \sin(m) \, I & \cos(m) \, T_+
\end{pmatrix},
\label{eq:u_dca}
\end{equation}
where \( T_\pm = \text{QFT} \cdot e^{\mp ik} \cdot \text{QFT}^\dagger = \sum_x |x \pm 1\rangle \langle x| \) are translation operators that shift the particle’s position by \( \pm 1 \). This discrete-time model enforces causal, unitary evolution of a Dirac particle on a lattice, with local interactions restricted to nearest neighbors.

The discretized wavefunction is represented by a two-component field \( \psi_R \) and \( \psi_L \), corresponding to the right-handed and left-handed components of the internal space. The one-step update rule in the real space is then
\begin{align}
\begin{pmatrix}
\psi_R(x, t+1) \\ \psi_L(x, t+1)
\end{pmatrix}
=& \cos(m)\begin{pmatrix}
\psi_R(x+1, t)  \\ \psi_L(x-1, t)
\end{pmatrix} \nonumber \\
& - i \sin(m)
\begin{pmatrix}
\psi_L(x, t) \\ \psi_R(x, t)
\end{pmatrix}.
\label{eq:one_step}
\end{align}
This evolution clearly couples the internal and external spaces, laying the foundation for studying entanglement and relativistic effects within a quantum computational framework.

\subsection{\label{sec:SDCA}Split-Step Dirac Cellular Automata}

Standard DCA approaches the continuous Dirac equation as the lattice spacing approaches zero and the number of sites $N$ becomes large. Since $c\delta t=\delta x$, increasing the spatial resolution simultaneously improves the temporal resolution, thereby reducing both spatial discretization and Trotterization errors. Therefore, a refined DCA can be realized by increasing the number of lattice sites $N$.

On a finite spatial lattice, we may preserve $\delta x=1$ and split the evolution operator into $M$ fractional sub-steps to approach continuous-time Dirac dynamics, where $M$ is the temporal refinement factor and each sub-step satisfies $c\delta t=\delta x/M$. In this approach, we define the split-step Dirac cellular automata (SDCA) from Eq.~\eqref{eq:u_dca} as
\begin{equation}
U^{(M)}_{\mathrm{SDCA}} =
\begin{pmatrix}
\cos(\frac{m}{M}) T_{-}^{1/M} & -i\sin(\frac{m}{M})I \\
-i\sin(\frac{m}{M})I & \cos(\frac{m}{M})T_{+}^{1/M}
\end{pmatrix}^M,
\label{eq:sdca}
\end{equation}
where $T_{\pm}^{1/M}=\text{QFT}Q_{\pm}^{1/M}\text{QFT}^{\dagger}$. In this notation, the SDCA reduces to the standard DCA when $M=1$, i.e., $U^{(1)}_{\mathrm{SDCA}}=U_{\mathrm{DCA}}$.

Since the DCA is a second-order approximation with error of $\mathcal{O}(\delta t^3)$, the SDCA refines this by partitioning $\delta t$ into $M$ sub-steps. Then the local error per sub-step is $\mathcal{O}((\delta t/M)^3)$, the global approximation error for the full time step $\delta t$ scales as:
\begin{equation}
\epsilon_{\text{global}} = M \times \mathcal{O}\left( \left(\frac{\delta t}{M}\right)^3 \right) = \mathcal{O}\left( \frac{\delta t^3}{M^2} \right).
\end{equation}
Thus, for a fixed spatial lattice $\delta x$, increasing $M$ quadratically reduces the Trotterization error.

The update rule for each sub-step is given by
\begin{align}
\begin{pmatrix}
\psi_R(x, t + \tfrac{1}{M}) \\ \psi_L(x, t + \tfrac{1}{M})
\end{pmatrix}
&= \cos\left(\frac{m}{M}\right) \sum_q
\begin{pmatrix}
A_q^{(-)}(M)\, \psi_R(x + q, t) \\
A_q^{(+)}(M)\, \psi_L(x - q, t)
\end{pmatrix} \nonumber \\
&\quad - i\sin\left(\frac{m}{M}\right)
\begin{pmatrix}
\psi_L(x, t) \\ \psi_R(x, t)
\end{pmatrix},
\label{eq:single_sdca}
\end{align}
where the transition amplitude from a neighbor at distance $q$ to site $x$ is defined as:
\begin{equation}
A_q^{(\pm)}(M) = \frac{1}{N} \cdot \omega^{-(1/M \pm q)/2} \frac{\sin(\pi (1/M \pm q))}{\sin(\frac{\pi}{N} (1/M \pm q))} 
\label{eq:transition_amp}
\end{equation}
with $N$ denoting the number of discrete external space positions. Here, a positive (negative) $q$ indicates a neighbor on the right-hand (left-hand) side (see Appendix~\ref{APP:A}). Unlike the standard DCA, in which the update is restricted to nearest-neighbor sites, the fractional translation operators in SDCA induce long-range transition amplitudes in the position representation. We therefore use the term SDCA to denote a split-step extension of the DCA framework rather than a strictly nearest-neighbor cellular automaton.

The position distribution of the particle at time $t$ is given by
\begin{equation}
P(x,t)=|\langle x|\Psi(t) \rangle|^{2}=|\psi_{R}(x,t)|^{2}+|\psi_{L}(x,t)|^{2},
\label{eq:prob_x}
\end{equation}
where $|\Psi(t) \rangle$ is the quantum state of the particle. In Fig.~\ref{fig:SDCA_circuit}(a), we demonstrate the spacetime diagram generated by the standard DCA and the SDCA. Comparing Eq.~\eqref{eq:sdca} with Eq.~\eqref{eq:one_step}, we observe that the SDCA introduces self-loop transitions within the internal space and a wider transition range across the external space. Figure~\ref{fig:SDCA_circuit}(b) illustrates the transition of probability amplitudes of the SDCA in the full Hilbert space, where R and L denote the right- and left-handed components of the internal space, respectively. Solid arrows represent external transitions, dashed arrows indicate internal transitions, and dotted arrows highlight the additional self-loop transitions introduced by applying $Q_{\pm}^{1/M}$ with $M>1$.

On the other hand, the velocity of the particle is evaluated by the velocity operator, which is defined as $V = \partial H/\partial p = -\sigma_z \otimes I_p$. The velocity expectation value is then given by:
\begin{eqnarray}
\langle V \rangle
    &=& -\mathrm{Tr}(\rho_c \sigma_z) \nonumber\\
    &=& \sum_{x} \left( |\psi_{R}(x,t)|^{2} - |\psi_{L}(x,t)|^{2} \right),
\label{eq:velocity_operator}
\end{eqnarray}
where
\begin{equation}
\rho_c = \sum_x
\begin{pmatrix}
|\psi_L(x,t)|^2 & \psi_L(x,t)\psi_R^*(x,t) \\
\psi_R(x,t)\psi_L^*(x,t) & |\psi_R(x,t)|^2
\end{pmatrix}
\label{eq:reduced_density_matrix}
\end{equation}
denotes the reduced density matrix of the internal space, obtained by taking the partial trace with respect to the external position space $\mathcal{H}_p$. Equation~\eqref{eq:velocity_operator} reveals that the velocity of the particle depends strictly on the probability difference between the right- and left-handed spinor components. In Fig.~\ref{fig:SDCA_circuit}(a), we show the time evolution of the velocity expectation value, demonstrating that the SDCA captures velocity oscillations, whereas the standard DCA does not.

\subsection{\label{sec:entanglement}Quantum Entanglement}

The Dirac dynamics implies that the internal and external space become entangled. We quantify the strength of entanglement by the von Neumann entropy:
\begin{equation}
S = -\mathrm{Tr}(\rho_c \log_2 \rho_c).
\label{eq:entropy}
\end{equation}

Since the internal space encodes the velocity and the external space encodes the position, the entropy \( S \) captures the physical correlations between the motion of the particle and its spin state. Diagonalizing \( \rho_c \) yields (see Appendix~\ref{APP:B})
\begin{equation}
S = -(\lambda_+ \log_2 \lambda_+ + \lambda_- \log_2 \lambda_-),
\label{eq:entropy_Dirac}
\end{equation}
where the eigenvalues of \( \rho_c \) are
\begin{equation}
\lambda_{\pm} = \frac{1}{2} \left(1 \pm \sqrt{ \langle \sigma_z \rangle_c^2 + \langle \sigma_x \rangle_c^2 + \langle \sigma_y \rangle_c^2 } \right).
\label{eq:lambda}
\end{equation}
Here, Eq.~\eqref{eq:lambda} shows that the entropy depends entirely on Pauli expectation values in the internal space. In particular, \( \langle \sigma_z \rangle_c = -\langle V \rangle \), while the others can be measured by basis changes:
\[
\langle \sigma_x \rangle_c = \langle \text{H} \sigma_z \text{H} \rangle_c, \quad
\langle \sigma_y \rangle_c = \langle R_x(\pi/2) \sigma_z R_x(-\pi/2) \rangle_c,
\]
where $\text{H}$ is the Hadamard gate and $R_{x}(\pi/2)$ is a single-qubit rotation gate about the $x$-axis.

\subsection{\label{sec:SDCA_circuit}The Circuit Implementation of SDCA}

The corresponding SDCA quantum circuits with refinement factor $M$ are shown in Fig.~\ref{fig:SDCA_circuit}(c), with the detailed circuits for the fractional shift operators $Q_{\pm}^{1/M}$ presented in Figs.~\ref{fig:SDCA_circuit}(d) and (e). The position distribution of the particle is obtained by measuring the external space, whereas the velocity expectation value can be determined solely by measuring the internal space. 

\begin{figure*}[ht]
\centering
\includegraphics[width=\linewidth]{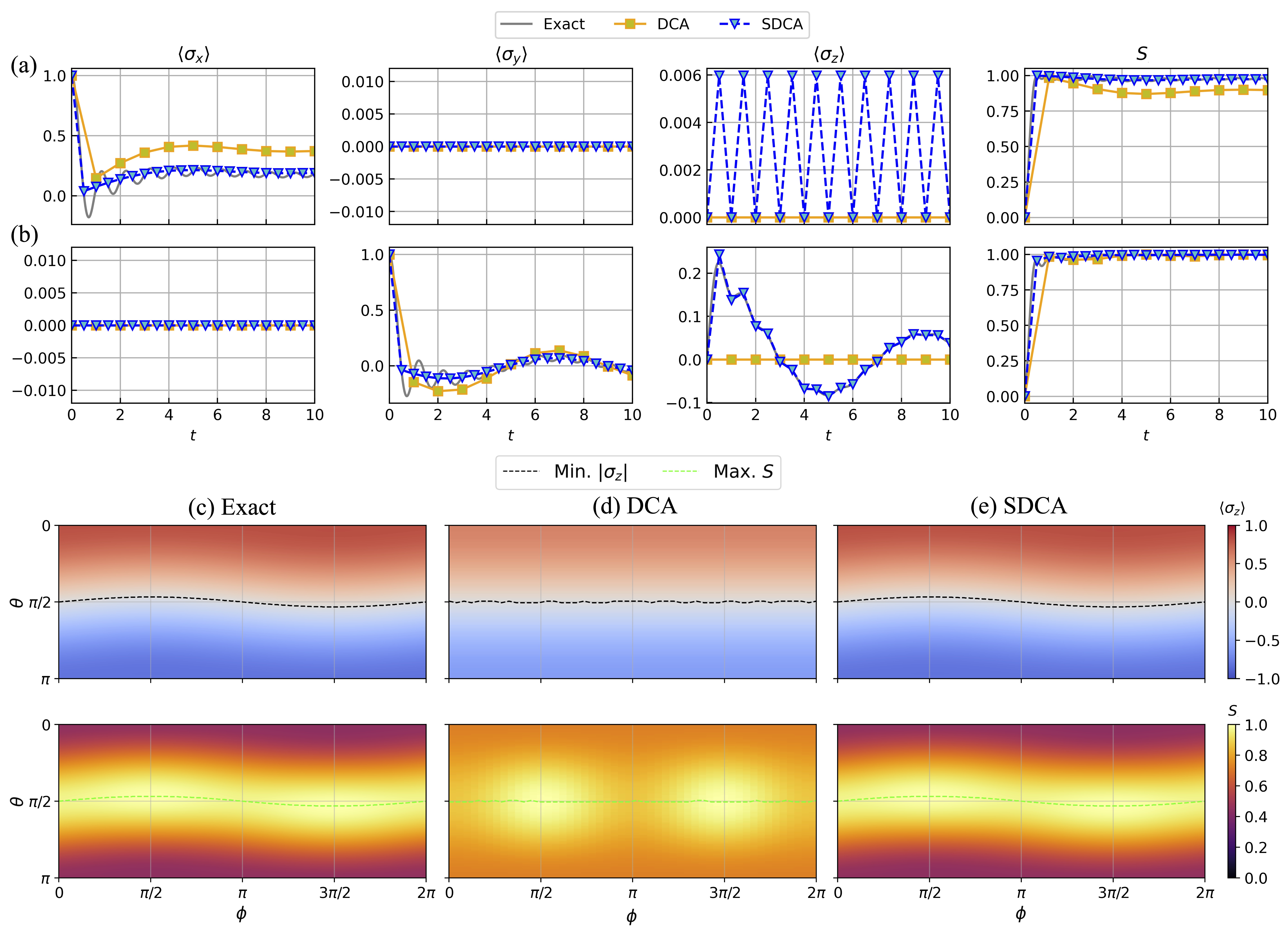}
\caption{Comparison of Dirac dynamics simulated using the DCA and the SDCA ($M=2$) against the exact continuous-time limit of a highly localized state $|\Psi(0)\rangle = (\cos(\theta/2)|0\rangle_{c} + e^{i\phi}\sin(\theta/2)|1\rangle_{c}) \otimes |x=0\rangle$ with mass $m=\pi/8$ on a finite spatial lattice. Top panels: Time evolution of the Pauli matrix expectation values ($\langle \sigma_x \rangle$, $\langle \sigma_y \rangle$, $\langle \sigma_z \rangle$) and the entanglement entropy $S$ with (a) $\theta=\pi/2$, $\phi=0$, and (b) $\theta=\pi/2$, $\phi=\pi/2$. Bottom panels: Heatmaps displaying $\langle \sigma_z \rangle$ and $S$ at time step $t = 5$ as a function of the initial state angles $\theta$ and $\phi$ for the (c) exact continuous limit, (d) DCA, and (e) SDCA.}
\label{fig:time_evolution} 
\end{figure*}

Importantly, in the momentum representation, the time-evolution operator is realized by parallel controlled-phase (CP) gates, requiring a star topology for qubit connectivity, as shown in Fig.~\ref{fig:time_evolution}(e). In this implementation, the expectation values of the internal space can be efficiently evaluated if the initial state of the particle is directly prepared in the momentum representation. This strategy allows the QFT and inverse QFT operations to be entirely omitted, reducing the circuit depth and gate overhead.

\section{\label{sec:dynamics} Dynamics Simulation Results}
One of the most iconic features of Dirac time evolution is Zitterbewegung, a relativistic quantum effect in which a free Dirac particle exhibits rapid oscillatory motion. This phenomenon arises because the velocity observable does not commute with the Dirac Hamiltonian $H$ (i.e., $[V, H] \neq 0$). Physically, this stems from the interference between positive and negative energy states (particles and antiparticles).

To demonstrate the effect of step splitting, we simulate the time evolution of a particle with mass $m = \pi/8$, initialized in a highly localized state:
\begin{equation}
    |\Psi(0)\rangle = \left(\cos(\theta/2)|0\rangle_{c} + e^{i\phi}\sin(\theta/2)|1\rangle_{c}\right) \otimes |x=0\rangle.
    \label{eq:initial_state}
\end{equation}
Figures~\ref{fig:time_evolution}(a) and (b) display the simulated dynamics for two specific initial states with equal left- and right-handed components ($\theta=\pi/2$), but distinct relative phases ($\phi=0$ and $\phi=\pi/2$). We observe that the two-step SDCA ($M=2$) exhibits Zitterbewegung oscillations that closely track the continuous-time limit, whereas the standard DCA completely suppresses them. The introduction of fractional sub-steps in the SDCA facilitates self-loop transitions in the internal space and broadens the transition range across the external space, generating physical interference effects that are absent in the rigid single-step model.

Furthermore, we analyze the dependence of these observables on the initial state angles $\theta$ and $\phi$ at $t=5$ (see Figs.~\ref{fig:time_evolution}(c)-(e)). We notice that the standard DCA yields a zero contour strictly fixed at $\theta=\pi/2$. In contrast, the zero contour of the SDCA matches the continuous-time limit, manifesting as a function of $\phi$. 

The time evolution of the expectation value $\langle \sigma_z \rangle_t$ for a state with equal left- and right-handed components is mathematically given by (see Appendix~\ref{APP:D}):
\begin{eqnarray}
\langle \sigma_{z} \rangle_t
&=& \frac{2\sin{(m)}}{N} \left[ \sin{(\phi)} \sum_{k} F_{m}(k,t) \right. \nonumber\\
 & & \left. - \cos{(\phi)} \cos{(m)}\sum_{k} G_{m}(k,t) \right],
\label{eq:balance_expval_z}
\end{eqnarray}
where $F_{m}(k,t)=f_{m}(k,t)\cos(\omega_k t)$ is an even function of $k$, and $G_{m}(k,t)=f_{m}^{2}(k,t)\sin{(k)}$ is an odd function of $k$. 

\begin{figure*}[ht]
\centering
\includegraphics[width=\linewidth]{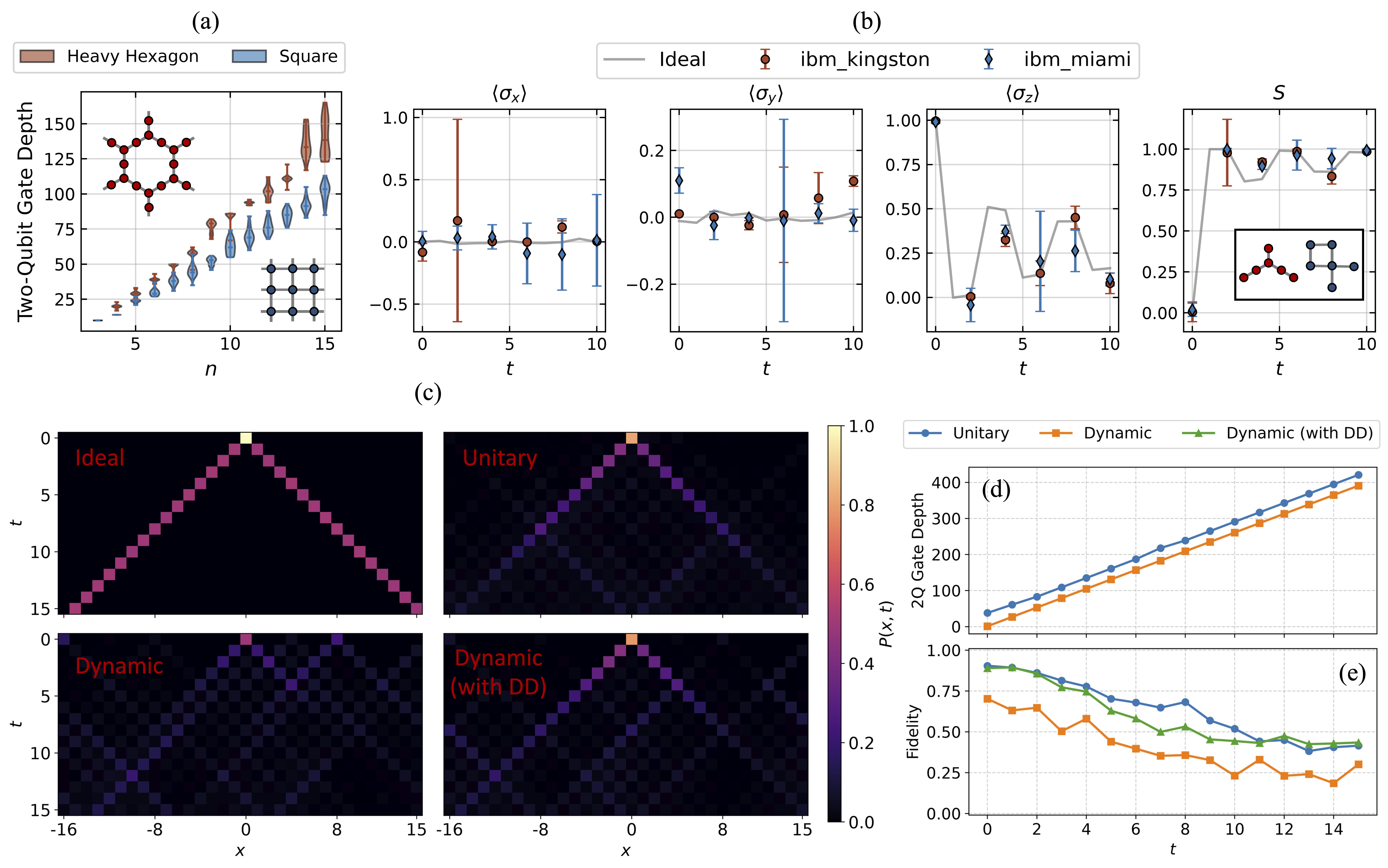}
\caption{
Experimental results of the DCA for a highly localized initial state.
    \textbf{Top Panel:} Comparison of results executed on heavy hexagon (\textit{ibm\_kingston}) and square (\textit{ibm\_miami}) lattice backends. 
    (a) Two-qubit gate depth scaling for single-step DCA circuits as a function of the external space size $n$. Insets illustrate the physical connectivity graphs of the respective hardware topologies. 
    (b) Time evolution of the three Pauli expectation values ($\langle \sigma_x \rangle$, $\langle \sigma_y \rangle$, $\langle \sigma_z \rangle$) and the entanglement entropy ($S$) for the particle state $|\Psi(0)\rangle = |0\rangle_{c} \otimes |x=0\rangle$ with mass $m=\pi/2$. The insets show the specific physical qubit mappings on the devices. 
    \textbf{Bottom Panel:} Performance comparison of standard unitary and dynamic circuit implementations on heavy hexagon (\textit{ibm\_pittsburgh}). 
    (c) Spacetime probability diagrams, $P(x,t)$, of a massless particle realized by the ideal theoretical model, a standard unitary circuit, a dynamic circuit, and a dynamic circuit with dynamical decoupling (DD). 
    (d) Two-qubit gate depth scaling versus time step $t$. 
    (e) Fidelity of the experimental probability distributions over time step $t$.}
\label{fig:real_device_DCA} 
\end{figure*}

In the standard DCA, the suppression of Zitterbewegung arises from an artificial phase-matching condition introduced by the rigid coupling between spatial ($\delta x$) and temporal ($\delta t$) resolutions ($\delta x = c\delta t$). This forces the amplitude to move exactly to adjacent lattice sites at each discrete time step. Because $F_{m}(k+\pi,t)=-F_{m}(k,t)$ and $G_{m}(k+\pi,t)=-G_{m}(k,t)$, the interference cross-terms perfectly cancel out when summed over the discrete Brillouin zone. Consequently, the entropy $S$ peaks at $\theta=\pi/2$ and exhibits this same artificial symmetry.

In contrast, the SDCA model mitigates this rigid coupling by splitting the time-evolution operator into $M$ sub-steps. The phase $\phi$ of the initial state physically interferes with the fractional kinetic phase introduced by the split-step operators $Q^{1/M}_{\pm}$. As a result, the peak of the entropy $S$ and the zero-value contour of $\langle \sigma_z \rangle_t$ correctly become functions of $\phi$.

It is worth noting the specific case of $\phi=0$. According to Eq.~\eqref{eq:balance_expval_z}, the first term vanishes entirely, leaving only the sum over the odd function $G_m(k,t)$. On a perfectly symmetric discrete grid, this integral evaluates exactly to zero. However, standard discrete momentum grids for QFT circuits of size $N=2^n$ are geometrically asymmetric; they include the negative boundary $-\pi$ but lack the positive boundary $+\pi$. When the fractional operator applies the complex phase $e^{-i\pi/M}$ to this unpaired boundary mode, it artificially breaks the $k \leftrightarrow -k$ parity symmetry of the grid. This discrete lattice artifact manifests as the small anomalous oscillation seen for SDCA in Fig.~\ref{fig:time_evolution}(a).

\section{\label{sec:NISQ_devices}Implementation on NISQ Devices}
\subsection{Hardware Topology Comparison}

The restricted qubit connectivity of real quantum hardware directly inflates the two-qubit gate depth of the transpiled circuits due to the necessary insertion of SWAP gates. Furthermore, two-qubit gates typically introduce higher errors than single-qubit gates, making their minimization crucial. 

For the hardware topology comparison presented in the top panel of Fig.~\ref{fig:real_device_DCA}, we operate in the momentum representation and only measure the internal space. Consequently, the QFT is omitted, and the qubits in the external space are initialized using Hadamard gates. This creates a uniform superposition in the momentum representation, which corresponds to a highly localized state ($|\Psi(0)\rangle =|0\rangle_c \otimes |x=0\rangle$) in the position representation. In Fig.~\ref{fig:real_device_DCA}(a), we demonstrate the two-qubit gate depth of DCA circuits transpiled onto heavy-hexagon (\textit{ibm\_kingston}) and square (\textit{ibm\_miami}) lattices; the higher qubit connectivity of the square lattice naturally results in lower SWAP gate overhead.

\begin{figure*}[t]
\includegraphics[width=\linewidth]{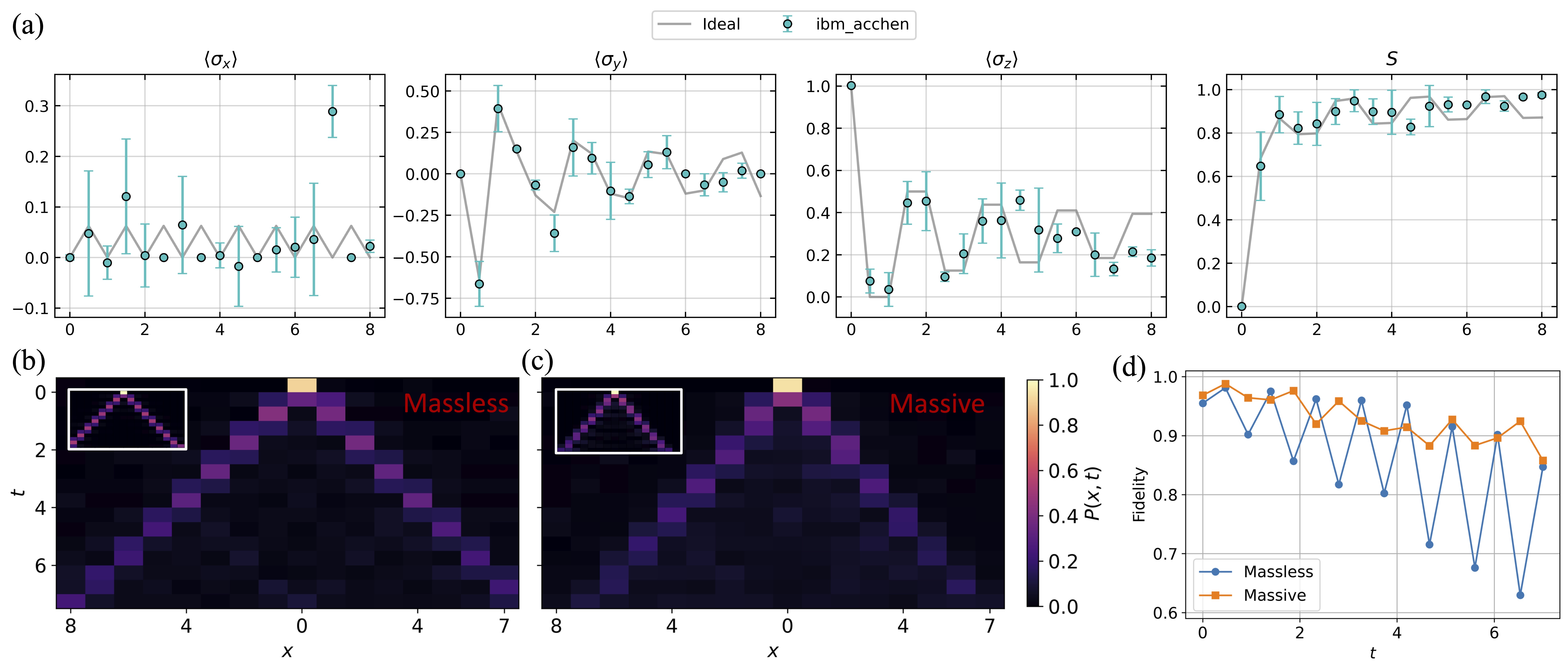}
\caption{Experimental results of the SDCA for a highly localized initial state executed on the \textit{ibm\_aachen} quantum processor. 
    (a) Time evolution of the internal space Pauli expectation values ($\langle\sigma_x\rangle$, $\langle\sigma_y\rangle$, $\langle\sigma_z\rangle$) and the corresponding entanglement entropy ($S$) for the initial state $|\Psi(0)\rangle = |0\rangle_{c}\otimes |x=0\rangle$ with mass $m=\pi/2$. (b) and (c) Experimental spacetime probability diagrams, $P(x,t)$ of the initial state $|\Psi(0)\rangle_c = \frac{1}{\sqrt{2}}(|0\rangle_{c}+|1\rangle_{c})\otimes |x=0\rangle$ for (b) a massless particle and (c) a massive particle with $m=\pi/4$. The insets display the corresponding ideal spacetime diagrams. 
    (d) Fidelity of the experimental probability distributions relative to the ideal distributions over time step $t$ for both the massless and massive cases.}
\label{fig:experiment} 
\end{figure*}

Figure~\ref{fig:real_device_DCA}(b) presents the experimental time evolution of the internal observables and entropy for a massive particle ($m=\pi/2$) with $n=5$. To mitigate the impact of hardware noise, we applied Twirled Readout Error Extinction (TREX) \cite{van2022model}, Dynamical Decoupling (DD) \cite{viola1999dynamical,jurcevic2021demonstration}, and Zero-Noise Extrapolation (ZNE) \cite{cai2021multi}. The results demonstrate a critical hardware trade-off. Although the \textit{ibm\_miami} (square lattice) backend yields a shallower two-qubit depth, the \textit{ibm\_kingston} (heavy-hexagon) backend produces superior experimental accuracy relative to the ideal simulation. This is directly attributed to the lower two-qubit gate and readout errors of the hardware as shown in Appendix~\ref{APP:E}.

\subsection{QFT Implementation Approaches}

To obtain the position probability distribution $P(x,t)$, the QFT must be implemented. In this experiment, we consider both unitary and dynamic implementation approaches. In the standard unitary implementation, the $n$-qubit QFT requires $\mathcal{O}(n^2)$ CP gates with all-to-all connectivity between the $n$ qubits. Conversely, the dynamic implementation combines the QFT with the final measurement. This technique replaces the CP gates with mid-circuit measurements (MCMs) paired with single-qubit classical feed-forward (FF) phase gates. Therefore, the dynamic implementation eliminates the two-qubit CP gates required by the unitary QFT and replaces them with $\mathcal{O}(n)$ MCMs and classical FF operations. (see Appendix~\ref{APP:D}).

In the bottom panel of Fig.~\ref{fig:real_device_DCA}, we evaluate different QFT implementations for a massless ($m=0$) particle as a benchmark. Figure~\ref{fig:real_device_DCA}(c) shows the spacetime diagrams generated by the ideal theoretical model, a standard unitary circuit, a dynamic circuit, and a dynamic circuit enhanced with DD. As shown in Fig.~\ref{fig:real_device_DCA}(d), the dynamic QFT implementation reduces the two-qubit gate depth scaling over time compared to the unitary approach. However, although the dynamic approach reduces the overall circuit depth, it suffers from measurement error accumulation and limited $T_1/T_2$ relaxation times. Consequently, even with the application of DD, the fidelity of the dynamic implementation reaches but does not surpass that of the unitary circuit, as demonstrated in Fig.~\ref{fig:real_device_DCA}(e).

\subsection{Dynamics of the SDCA}
In Fig.~\ref{fig:experiment}, we implemented the two-substep SDCA on the \textit{ibm\_aachen} backend. Due to the finer temporal resolution, the SDCA inherently demands a deeper quantum circuit. As shown in Fig.~\ref{fig:experiment}(a), the experimental results for the internal space observables demonstrate that the hardware closely tracks the interference oscillations, particularly during the early time steps ($t \le 4$). Furthermore, the experimental entanglement entropy $S$ rapidly saturates and matches the ideal curve, confirming that the proper quantum correlations between the internal and external spaces are accurately generated. However, as the time evolution progresses ($t > 4$), the accumulated gate errors from the deep circuit begin to manifest.

Figures~\ref{fig:experiment}(b) and (c) display the experimental spacetime diagrams for a massless particle and a massive particle ($m=\pi/4$) via unitary QFT, respectively. Despite the deep circuits required for the SDCA, the experimental distributions capture the physical light-cone structure and interference patterns seen in the ideal theoretical insets. Figure~\ref{fig:experiment}(d) illustrates the corresponding fidelity of these distributions over time. While the fidelity naturally decays due to the accumulation of gate errors over the deep circuits, the fidelity of the SDCA exhibits an oscillatory behavior. For a massless particle, the position probability distribution is wider at odd time steps, which obscures the physical dispersion due to hardware noise. Similarly, a massive particle exhibits a wider spatial distribution than a massless particle, which results in a higher experimental fidelity.

\section{\label{sec:conclusion} Conclusion}

In this study, we extend the standard DCA to the SDCA to simulate Dirac dynamics with a finite spatial lattice. Standard DCA models are fundamentally limited by a rigid spatio-temporal resolution coupling ($c\delta t = \delta x $). In contrast, the SDCA introduces self-loop transitions in the internal space and a wider transition range across the external space via its split-step evolution. Furthermore, the rigid coupling of the standard DCA manifests as an artificial phase-matching symmetry across the Brillouin zone. For the balanced localized initial states considered here, Zitterbewegung is exactly suppressed by this symmetry. By Trotterizing the time-evolution operator into $M$ fractional sub-steps in the momentum representation, the SDCA effectively breaks this artificial symmetry. This allows the model to capture the velocity oscillations and entanglement dynamics of the continuous-time limit.

We also provided a hardware-level optimization analysis of the implementation of SDCA. The comparison between different hardware topologies demonstrates that higher connectivity can reduce SWAP overhead but does not necessarily lead to higher fidelity, as gate and measurement errors remain important. In addition, to address the deep circuits required for the position distribution measurement, we explored dynamic circuit implementations of the QFT. By replacing  CP gates with MCMs and classical FF operations, the dynamic QFT curtails gate depth scaling, though its current fidelity remains bottlenecked by measurement and relaxation times.

Ultimately, the SDCA improves the temporal resolution of relativistic quantum simulations on fixed spatial registers. Furthermore, its sub-step architecture naturally supports adaptive Trotterization (ADA Trotter) for simulating time-dependent or interacting systems. By operating strictly without spatial overhead, the SDCA provides a practical blueprint for using near-term quantum processors to probe fundamental physics.

\begin{acknowledgments}
W.T.W, and C.R.C acknowledge funding from the National Science and Technology Council grant No. NSTC 113-2112-M-033-011 (Taiwan). P.M.H. is supported in part by the Ministry of Science and Technology, R.O.C. (NSTC 112-2112-M-002 -024 -MY3, NSTC 113-2112-M-002 -040 -MY2). We are also supported in part by National Taiwan University.
\end{acknowledgments}

\appendix

\section{The Transition Amplitude in Split-step Dirac Cellular Automata}
\label{APP:A}

Considering that the quantum system evolves in a finite Hilbert space with $N$ discrete sites, the corresponding phase space is a quantized toroidal phase space with unit $\Delta k = 2\pi/N$. The QFT maps the state from position representation to momentum representation and has the matrix form:
\begin{equation}
\mathrm{QFT} = 
\frac{1}{\sqrt{N}}\begin{pmatrix}
1 & 1 & 1 & \cdots & 1 \\
1 & \omega & \omega^2 & \cdots & \omega^{N-1} \\
1 & \omega^2 & \omega^4 & \cdots & \omega^{2(N-1)} \\
\vdots & \vdots & \vdots & \ddots & \vdots \\
1 & \omega^{N-1} & \omega^{2(N-1)} & \cdots & \omega^{(N-1)(N-1)}
\end{pmatrix}_{N \times N},
\label{eq:QFT}
\end{equation}
where $\omega=e^{i\Delta k}=e^{2\pi i/N}$.  To analyze the position representation transition amplitudes in the SDCA model, we transform $Q_{+}^{1/M}$
\begin{equation}
Q_{-}^{1/M} = 
\begin{pmatrix}
1 & 0 & \cdots & 0 & 0 & \cdots & 0 \\
0 & \omega^{1/M} & \cdots & 0 & 0 & \cdots & 0 \\
\vdots & \vdots & \ddots & \vdots & \vdots & \ddots & \vdots \\
0 & 0 & \cdots & \omega^{\frac{N/2-1}{M}} & 0 & \cdots & 0 \\
0 & 0 & \cdots & 0 & \omega^{-\frac{N}{2M}} & \cdots & 0 \\
\vdots & \vdots & \ddots & \vdots & \vdots & \ddots & \vdots \\
0 & 0 & \cdots & 0 & 0 & \cdots & \omega^{-\frac{1}{M}}
\end{pmatrix}_{N \times N}.
\label{eq: fractional_Q}
\end{equation}
back to the real space by
\begin{equation}
 T_{-}^{1/M}= \frac{1}{N} \sum_{j=-N/2}^{N/2-1}\omega^{j/M}
\begin{pmatrix}
1 & \omega^{-j} & \omega^{-2j} & \cdots & \omega^{j} \\
\omega^{j} & 1 & \omega^{-j} & \ddots & \vdots \\
\omega^{2j} & \omega^{j} & 1 & \ddots & \vdots \\
\vdots & \ddots & \ddots & \ddots & \vdots \\
\omega^{-j} & \cdots & \cdots & \cdots & 1
\end{pmatrix}_{N \times N}.
\label{eq:Q_in_real}
\end{equation}
where $T_{+}^{1/M}=\mathrm{QFT}Q_{+}^{1/M}\mathrm{QFT}^{\dagger}$.
This operator governs the probability amplitude for the particle to move from one site to another after a split-step evolution. 

By defining $z=(1/M - q)$, the transition amplitude from any arbitrary site to its $q$-th neighbor is
\begin{align}
A_q^{(-)}(M) 
\nonumber
&= \frac{1}{N} \sum_{j=-N/2}^{N/2-1} \omega^{jz} \\
\nonumber
&= \frac{1}{N} \cdot (\omega^{(-N/2)z}+\omega^{(-N/2+1)z}+\dots+\omega^{(N/2-1)z}) \\
\nonumber
&= \frac{1}{N} \cdot \frac{\omega^{-\frac{N}{2}z}(1 - \omega^{Nz})}{1 - \omega^{z}} \\
\nonumber
&= \frac{1}{N} \cdot \frac{(\omega^{-\frac{N}{2}z} - \omega^{\frac{N}{2}z})}{\omega^{z/2}(\omega^{-z/2} - \omega^{z/2})} \\
\nonumber 
&= \frac{1}{N} \cdot \omega^{-z/2} \frac{\sin(\pi z)}{\sin(\pi z/N)} \\
&= \frac{1}{N} \cdot \omega^{-(1/M - q)/2} \frac{\sin(\pi (1/M - q))}{\sin(\frac{\pi}{N} (1/M - q))},  
\label{eq:A_n}
\end{align}
where $z=(1/M - q)$ and $q>0$ ($q<0$) denotes transitions to the right (left).
Similarly, for the backward operator $Q_+^{(1/M)}$, the transition amplitude to its $q$-th neighbor is:
\begin{equation}
A_q^{(+)}(M) = \frac{1}{N} \cdot \omega^{-(1/M + q)/2} \frac{\sin(\pi (1/M + q))}{\sin(\frac{\pi}{N} (1/M + q))} 
\label{eq:A_p}
\end{equation}
These results demonstrate that step-splitting (via $M$) spreads the transition amplitude to distant sites, leading to wider propagation in position representation. The expressions also reveal how the discrete phase shifts give rise to complex interference patterns as a function of $q$ and $M$.

\section{Entanglement and Velocity from Reduced Density Matrix}
\label{APP:B}

To investigate the connection between quantum entanglement and the velocity of a Dirac particle, we consider the reduced density matrix of the internal space, obtained by tracing out the external space from the full system’s pure-state density matrix $\rho$. The reduced density matrix $\rho_c = \mathrm{Tr}_p(\rho)$ is given by:
\begin{equation}
\rho_c = \sum_x 
\begin{pmatrix}
|\psi_L(x)|^2 & \psi_L(x) \psi_R^*(x) \\
\psi_R(x) \psi_L^*(x) & |\psi_R(x)|^2
\end{pmatrix}
\label{eq:rho_c}
\end{equation}
The von Neumann entropy of the reduced density matrix quantifies the entanglement between internal and external spaces:
\begin{equation}
S = -\mathrm{Tr}(\rho_c \log_2 \rho_c) = -\lambda_+ \log_2 \lambda_+ - \lambda_- \log_2 \lambda_-,
\end{equation}
where $\lambda_\pm$ are the eigenvalues of $\rho_c$, given by:
\begin{align}
\lambda_\pm &= \frac{1}{2} \left( 1 \pm \sqrt{1 - 4(\Sigma_{L^2} \Sigma_{R^2} - |\Sigma_{LR}|^2)} \right) \notag \\
&= \frac{1}{2} \left( 1 \pm \sqrt{\langle \sigma_z \rangle_c^2 + \langle \sigma_x \rangle_c^2 + \langle \sigma_y \rangle_c^2} \right)
\end{align}
where $\Sigma_{L^2} = \sum_x |\psi_L(x)|^2$, $\Sigma_{R^2} = \sum_x |\psi_R(x)|^2$, and $\Sigma_{LR} = \sum_x \psi_L(x) \psi_R^*(x)$.
In the final expression, we have used the Bloch sphere representation of $\rho_c$, where the Bloch vector components are given by:
\begin{align}
    \langle \sigma_z \rangle_c &= \mathrm{Tr}(\rho_c \sigma_z) = \Sigma_{L^2} - \Sigma_{R^2} \\
\langle \sigma_x \rangle_c &= \mathrm{Tr}(\rho_c \sigma_x) = \Sigma_{LR} + \Sigma_{LR}^* \\
\langle \sigma_y \rangle_c &= \mathrm{Tr}(\rho_c \sigma_y) = i(\Sigma_{LR} - \Sigma_{LR}^*)
\end{align}
This establishes that the entanglement entropy depends on the magnitude of the Bloch vector, which includes the expectation value of $\sigma_z$, and hence is directly related to the particle’s velocity.

\begin{figure}[t]
\includegraphics[width=\linewidth]{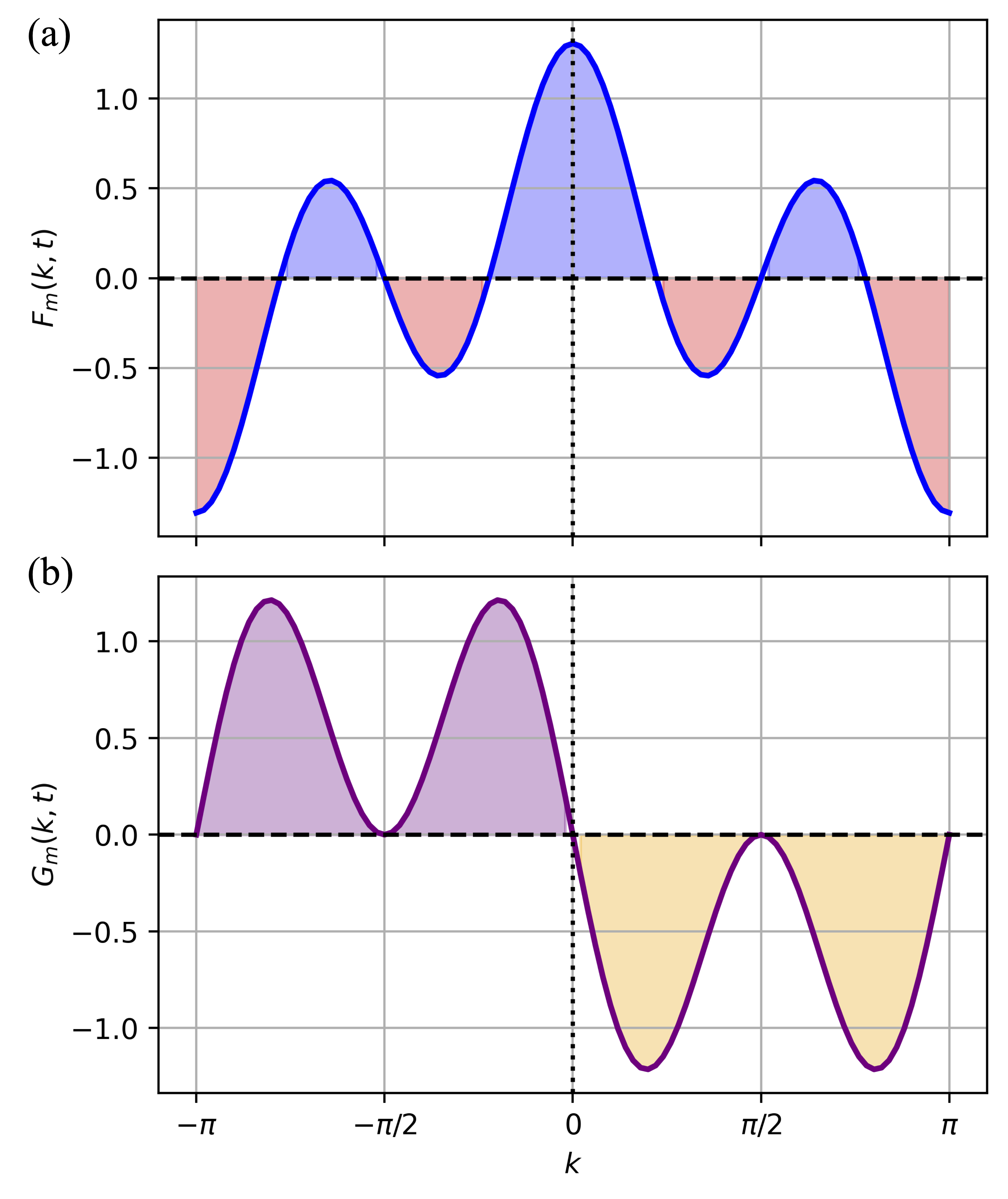}
\caption{The time-dependent functions (a) $F_{m}(k,t)$ and (b) $G_{m}(k,t)$ plotted over the Brillouin zone with $t=2$. Both functions exhibit anti-symmetry under a $k \to k+\pi$ momentum shift.}
\label{fig: F_and_G_function} 
\end{figure}

\begin{figure}[t]
\includegraphics[width=\linewidth]{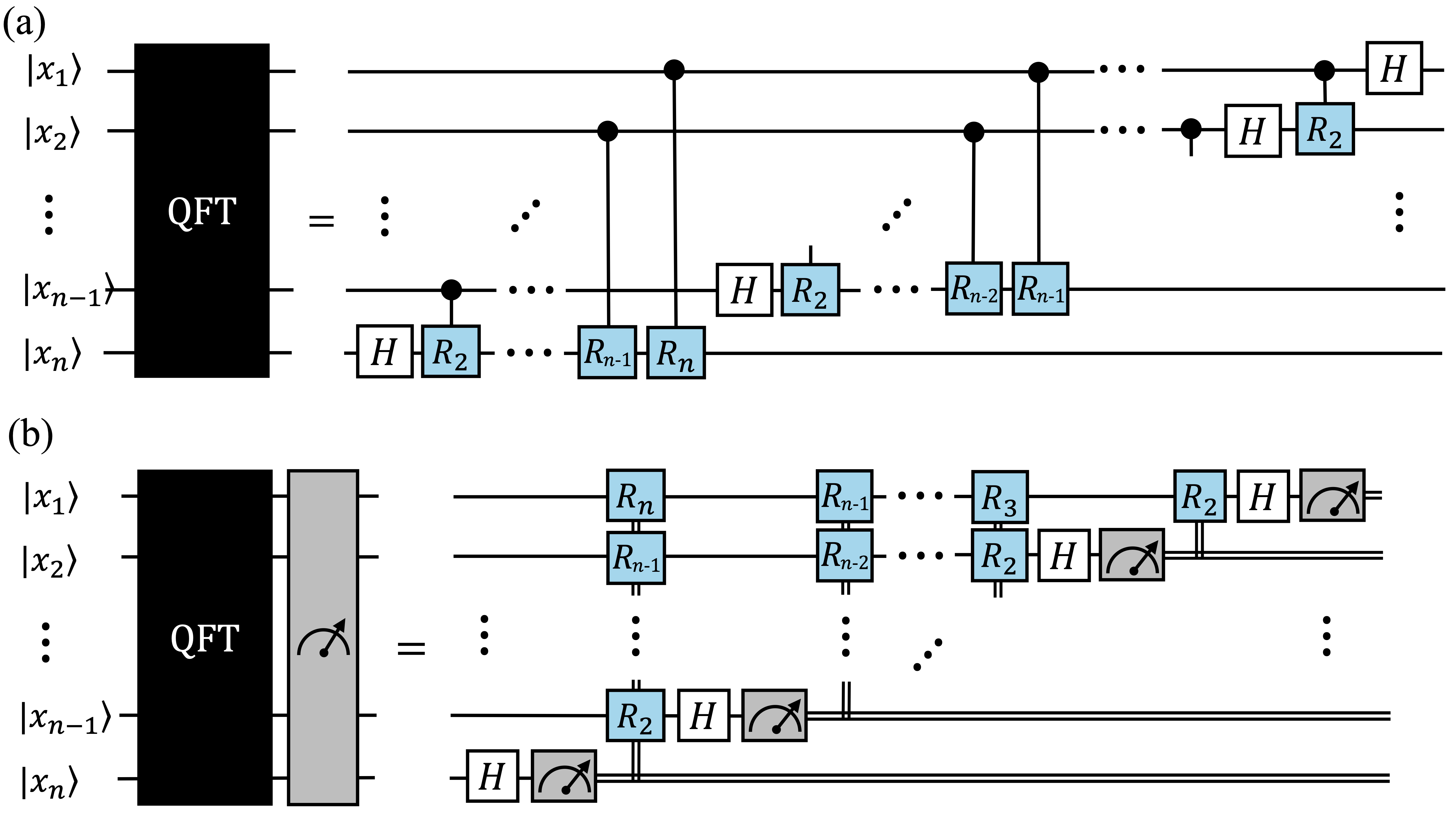}
\caption{The quantum Fourier transform implementation via (a) a standard unitary circuit and (b) a dynamic circuit, where
$R_{k}=\begin{pmatrix} 1 & 0 \\ 0 & e^{2\pi i/2^{k}} \end{pmatrix}$.}
\label{fig:SQFT} 
\end{figure}

\begin{figure}[ht]
\centering
\includegraphics[width=\linewidth]{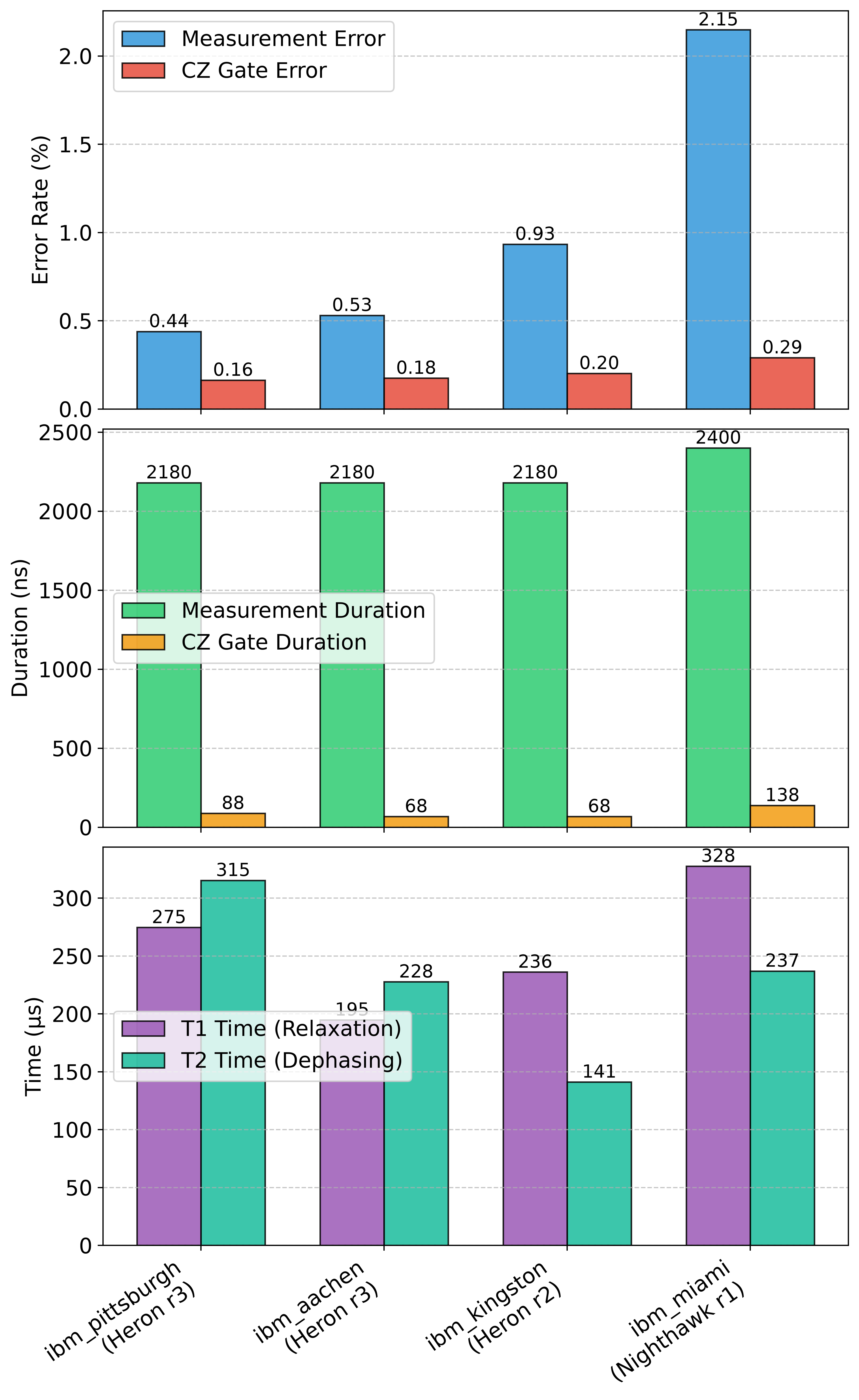}
\caption{Hardware calibration data for the IBM Quantum processors used in this work, detailing median measurement and CZ gate error rates (top), measurement and CZ gate durations (middle), and relaxation ($T_1$) and dephasing ($T_2$) times (bottom).}
\label{fig:calibration_data}
\end{figure}

\section{Balanced Propagation of DCA}
\label{APP:C}

Let the single-step DCA operator in the momentum representation $U(k)$ be defined as:
\begin{equation}
U(k) = 
\begin{pmatrix} 
\cos(m)e^{ik} & -i\sin(m) \\ 
-i\sin(m) & \cos(m)e^{-ik} 
\end{pmatrix}.
\end{equation}
Using Euler's formula $e^{\pm ik} = \cos k \pm i \sin k$, and the dispersion relation $\cos(\omega_k)=\cos(m)\cos(k)$, we can decompose the operator into the Pauli basis:
\begin{equation}
U(k) = e^{-i \omega_k (\hat{n} \cdot \vec{\sigma})} = \cos(\omega_k) I - i \sin(\omega_k) (\hat{n} \cdot \vec{\sigma}),
\end{equation}
where the eigenfrequency $\omega_k$ satisfies $\sin^2(\omega_k)=1-\cos^2(m)\cos^2(k)$, and the rotation axis is given by $\hat{n} = \frac{1}{\sin(\omega_k)}(\sin(m), 0, -\cos(m)\sin(k))$.

Consequently, the evolution operator for $t$ steps corresponds to a rotation by an angle $t\omega_k$:
\begin{equation}
U^t(k) = \cos(\omega_k t) I - i \sin(\omega_k t) (\hat{n} \cdot \vec{\sigma}).
\end{equation}
Substituting the expression for $\hat{n} \cdot \vec{\sigma}$ back into the matrix form yields:
\begin{align}
U^t(k) &= C_{\omega_k t} I + i f_{m}(k,t) \left[ C_mS_k \sigma_z - S_m \sigma_x \right],
\end{align}
where we use the shorthand where $f_{m}(k,t) = \frac{\sin(\omega_{k}t)}{\sin(\omega_{k})}$, $C_{\omega_k t} = \cos(\omega_k t)$, $C_m = \cos(m)$, $S_m = \sin(m)$, and $S_k = \sin(k)$. Expanding the Pauli matrices returns the explicit time-evolved matrix:
\begin{equation}
U^t = 
\begin{pmatrix}
C_{\omega_k t} + i f_{m}(k,t) C_m S_k & -i f_{m}(k,t) S_m \\
-i f_{m}(k,t) S_m & C_{\omega_k t} - i f_{m}(k,t) C_m S_k
\end{pmatrix}.
\end{equation}

For the localized initial state defined in Eq.~\eqref{eq:initial_state} with $\theta=\pi/2$, the spinor in momentum representation is uniform:
\begin{equation}
\begin{pmatrix}
\psi_R(k, t=0) \\ \psi_L(k, t=0)
\end{pmatrix}
=\frac{1}{\sqrt{2N}}
\begin{pmatrix}
1 \\ e^{i\phi}
\end{pmatrix}.
\end{equation}
Then, the expectation value $\langle\sigma_{z}\rangle$ at time $t$ is calculated as the difference between the upper and lower components:
\begin{eqnarray}
\langle \sigma_{z} \rangle_t &=& \sum_{k} \left( |\psi_{R}(k,t)|^{2} - |\psi_{L}(k,t)|^{2} \right) \nonumber\\
&=& \frac{2}{N}\sum_{k} f_{m}(k,t)S_{m} \Big[ C_{\omega_k t}\sin{(\phi)} \nonumber\\
 & & - f_{m}(k,t)C_{m}S_{k}\cos{(\phi)} \Big] \nonumber\\
&=& \frac{2\sin{(m)}}{N} \left[ \sin{(\phi)} \sum_{k} F_{m}(k,t) \right. \nonumber\\
 & & \left. - \cos{(\phi)} \cos{(m)}\sum_{k} G_{m}(k,t) \right],
\end{eqnarray}
where $F_{m}(k,t)=f_{m}(k,t)\cos(\omega_k t)$ is an even function of $k$, and $G_{m}(k,t)=f_{m}^{2}(k,t)\sin{(k)}$ is an odd function of $k$, as shown in Fig.~\ref{fig: F_and_G_function}.

The vanishing of $\langle \sigma_{z} \rangle_t$ in the standard DCA arises from the anti-symmetry properties of the summand under the momentum shift $k \to k + \pi$. Specifically, because $\omega_{k+\pi} = \pi - \omega_k$ and time $t$ is an integer, the relevant functions rigorously satisfy the following relations for all $t$:
\begin{eqnarray}
F_{m}(k+\pi,t) &=& -F_{m}(k,t), \\
G_{m}(k+\pi,t) &=& -G_{m}(k,t).
\end{eqnarray}
These identities indicate that when summing over the $N=2^n$ discrete momentum values $k_j = \frac{2\pi j}{N}$ for $j \in [-N/2, N/2 - 1]$, the terms in the left half of the Brillouin zone perfectly cancel the corresponding shifted terms in the right half. This pairwise cancellation yields a total expectation value of $\langle \sigma_{z} \rangle_t=0$ at all time steps. Consequently, we rigorously conclude that the initial phase $\phi$ does not affect the balanced propagation of the standard DCA, and Zitterbewegung is entirely suppressed by this discrete phase-matching symmetry.

\section{Semi-classical Quantum Fourier Transformation}
\label{APP:D}

Figure~\ref{fig:SQFT}(a) illustrates the standard unitary circuit of the QFT for an $n$-qubit system. If the QFT is immediately followed by measurement (QFT+M), it can be realized using a dynamic circuit, as shown in Fig.~\ref{fig:SQFT}(b).

\section{Calibration Data of IBM Quantum Processors}
\label{APP:E}

For completeness, Fig.~\ref{fig:calibration_data} provides the calibration data for the IBM Quantum processors utilized in our experiments, including the median gate error rates, median gate and measurement durations, and median qubit coherence times ($T_1$ and $T_2$).


\bibliography{apssamp}

\end{document}